\documentclass[preprint,showpacs,preprintnumbers,amsmath,amssymb]{revtex4}

\usepackage{graphicx}
\usepackage{dcolumn}
\usepackage{bm}

\begin{document}

\preprint{}

\title{Two-Qubit 
Separabilities as Piecewise Continuous Functions of Maximal Concurrence}

\author{Paul B. Slater}%
\email{slater@kitp.ucsb.edu}
\affiliation{%
ISBER, University of California, Santa Barbara, CA 93106\\
}%
\date{\today}

\begin{abstract}
The generic real ($\beta=1$) and complex 
($\beta=2$) two-qubit states are
9-dimensional and 15-dimensional in nature, respectively. 
The {\it total} volumes of 
the spaces they occupy with respect to the Hilbert-Schmidt and Bures 
metrics are obtainable as special cases of formulas of \.Zyczkowski and
Sommers. We claim that if one could determine certain
metric-{\it independent} 3-dimensional ``{\it eigenvalue}-parameterized 
separability 
functions'' (EPSFs), $S_4^{(1,\beta)}(\lambda_1\ldots\lambda_4)$, 
then these formulas could be
readily modified so as to yield the Hilbert-Schmidt and 
Bures volumes occupied by only the {\it separable} two-qubit states 
(and hence associated separability {\it probabilities}).   
Motivated by analogous earlier analyses of 
``{\it diagonal-entry}-parameterized separability functions'',
we further explore the possibility that such 
3-dimensional EPSFs might, in turn, be expressible as {\it univariate}
functions of some special relevant variable--which we hypothesize 
to be  the
maximal {\it concurrence}
($0 \leq C \leq 1$) over spectral orbits. Extensive numerical results 
we obtain are rather closely supportive
of this hypothesis.  {\it Both} the real and complex estimated 
EPSFs  exhibit
clearly pronounced {\it jumps} of magnitude roughly $50\%$ at 
$C=\frac{1}{2}$, as well as a number of additional {\it matching} 
discontinuities.

\end{abstract}


\maketitle
\section{Introduction}
In a pair of major, skillful papers, using concepts of random matrix theory, 
\.Zyczkowski and Sommers were able to obtain exact formulas for the 
{\it total} volumes--both in terms of the Hilbert-Schmidt (HS) 
metric \cite{szHS} and Bures (minimal monotone) metric \cite{szBures}--of the $(N^2-1)$-dimensional convex set of $N \times N$ complex density matrices and the 
$((N^2+N-2)/2)$-dimensional convex set of $N \times N$ real density matrices, 
representing $N$-level quantum systems 
(cf. \cite{andai} \cite[secs. 14.3, 14.4]{ingemarkarol}). 
In two recent studies, we have been interested in the question of how to 
modify/truncate, in some natural manner (by multiplying certain
integrands by relevant functions), these formulas of
\.Zyczkowski and Sommers, so that they will yield not {\it total} volumes, but only the (lesser, strictly included) 
volumes occupied by the {\it separable/nonentangled} states 
\cite{slaterJGP2,slaterJMP2008} (cf. \cite{ZHSL}).
We will below report some interesting progress in this regard, in relation to the two-qubit ($N=4$) states.

To begin, we present 
two parallel formulas from 
\cite{szHS} and \cite{szBures} for certain generalized normalization
constants ($C_{N}^{(\alpha,\beta)})$ 
used in the total HS and Bures volume computations. 
(Some notation and formatting has been altered.)
For the HS case, we have \cite[eq. (4.1)]{szHS} 
(cf. \cite[eq. (14.35)]{ingemarkarol}),
\begin{equation} 
\frac{1}{C_{N(HS)}^{(\alpha,\beta)}}=
\int_0^{\infty}  \prod_{i=1}^{N}
{\rm d}\lambda_{i }
\delta(\sum_{i=1}^N \lambda_i -1) \prod_{i=1}^N \lambda_i^{\alpha-1}
\prod_{i<j} |\lambda_i-\lambda_j|^{\beta},
\label{constab}
\end{equation}
and for the Bures case \cite[eq. (3.19)]{szBures} 
(cf. \cite[eq. (14.46)]{ingemarkarol}),
\begin{equation} 
 \frac{1}{C_{N(Bures)}^{(\alpha,\beta)}}=\int_0^{\infty} \prod_{i=1}^{N}
\frac{{\rm d}\lambda_{i }}{\lambda_{i}^{1/2}}
\delta(\sum_{i=1}^N \lambda_i -1)
\left[\prod_{i<j}^{1...N} \frac{(\lambda_{i} -\lambda_{j})^2}{\lambda_{i}+\lambda_{j}} 
\right]^{\beta/2}\prod_{i ={1}}^N
\lambda_{i}^{\alpha-1}
 \label{CHall}\ .
\end{equation}
The $\lambda$'s are the $N$ (nonnegative) eigenvalues--constrained to 
sum to 1--of the corresponding $N \times N$ density 
matrices, while 
the parameter $\beta$ is a ``Dyson index'', with $\beta=1$ corresponding 
to the real case, and $\beta=2$, the complex case (and $\beta=4$, the 
quaternionic case, not explicitly discussed in \cite{szHS,szBures}). 
The parameter $\alpha$ will be
equal to 1 for the case--of immediate interest to us here--of 
generically {\it nondegenerate} density matrices.
\subsection{Objective}
Our goal, in overall terms, is to find metric-{\it independent}
(separability) functions,
\begin{equation} \label{goal}
S_N^{(\alpha,\beta)}(\lambda_1\ldots\lambda_N),
\end{equation}
which, if inserted into formulas (\ref{constab}) and (\ref{CHall}) 
under the integral signs, as simple 
multiplicative factors, will yield 
separable-rather than total--volumes when the resulting modified 
$C_{N}^{\alpha,\beta}$'s are employed in exactly the same 
auxiliary computations (involving flag manifolds) in
\cite{szHS} and \cite{szBures} as the $C_{N}^{\alpha,\beta}$'s given by 
(\ref{constab}) and (\ref{CHall}) 
were there. 
More specifically here, our numerical analyses will be 
restricted to the 
$N=4$ and $\beta=2$ (complex) and $\beta=1$ (real) cases. 

Our metric-independent goal is plausible for the following reason.
Precisely the same
preliminary integrations--respecting the separability
constraints--over the non-eigenvalue 
parameters [possibly, Euler angles \cite{tbs} \cite[App. I]{slaterJMP2008}] 
must be performed for both metrics before arriving at the stage at which
we must concern ourselves with 
the remaining integration over the eigenvalues and
the {\it differences} that are now clearly apparent  
between metrics in their corresponding 
measures over the simplex 
of eigenvalues. Although we are not able to explicitly/symbolically 
determine what the results of
these preliminary integrations might be (the computational 
challenges are certainly considerable), they must--whatever form 
they may take--obviously be the same
for both metrics in question. Our goal here is to understand--with 
the assistance of numerical methods--what 
functional forms these preliminary (12-dimensional in the complex 
case, and 9-dimensional in the real case) integrations yield.
\subsection{Maximal concurrence and absolute separability}
The further narrower specific 
focus of this study will be to explore the possibility
that there exists a functional relationship of the form,
\begin{equation}\label{ansatz}
S_4^{(1,\beta)}(\lambda_1\ldots\lambda_4) = \sigma^{(\beta)} 
(C(\lambda_1\ldots\lambda_4)),
\end{equation}
where $\sigma^{(\beta)}(x)$ are some unknown {\it univariate} ({\it one}-dimensional)
functions and
\begin{equation} \label{maxcon}
C(\lambda_1\ldots\lambda_4)= \max \{0,\lambda_1 -\lambda_3 -2 
\sqrt{\lambda_2 \lambda_4}\}, 
\hspace{.2in} \lambda_1 \geq \lambda_2 \geq \lambda_3 \geq \lambda_4,
\end{equation}
is the {\it maximal concurrence} over spectral orbits of 
two-qubit density matrices \cite[sec. VII]{roland2} \cite{ishi,ver}.

For two-qubit states, $C \in 
[0,1]$, with $C=0$ corresponding
to the {\it absolutely} separable states. That is, {\it no} 
density matrix with
$C=0$ can be nonseparable/entangled \cite{wkw}. 
(In a recent study, we were able to obtain {\it exact} expressions--involving
the tetrahedral dihedral angle $\cos ^{-1}\left(\frac{1}{3}\right)$--for 
the contributions to the Hilbert-Schmidt real and complex two-qubit volumes 
for those states with $C=0$, and to numerically estimate the Bures
counterparts
\cite[secs. III.B, III.C]{slaterJMP2008}. In numerical terms, the HS
{\it absolute} 
separability probability of generic complex two-qubit states is 
0.00365826, and the Bures counterpart, 0.000161792. The HS real 
analogue is 0.0348338.) The {\it concurrence} itself
is a widely used entanglement measure of bipartite mixed states 
\cite[eq. (15.26)]{ingemarkarol}. 
\subsection{Motivation}
Certainly part of 
our motivation for advancing the ansatz (\ref{ansatz}) was that an
analogous modeling of a {\it trivariate} function in terms of a {\it 
univariate} 
function was found to hold--making use of the {\it Bloore} 
(correlation coefficient) 
parameterization of density matrices \cite{bloore}--for 
{\it diagonal-entry}-parameterized separability
functions  \cite[eq. (6)]{slaterPRA2} \cite{slater833}. This led to 
substantial insights--and {\it exact} conjectures 
($\frac{8}{33}$ and $\frac{8}{17}$)--with regard to Hilbert-Schmidt 
(complex and real) two-qubit separability probabilities. 
(The Dyson indices $\beta$ played a central analytical role there, 
in relating real and complex [and quaternionic] results,
but not apparently--as far as we can perceive--in the analyses to
be presented below.)
\section{Numerics}
\subsection{Methodology}
\begin{figure}
\includegraphics{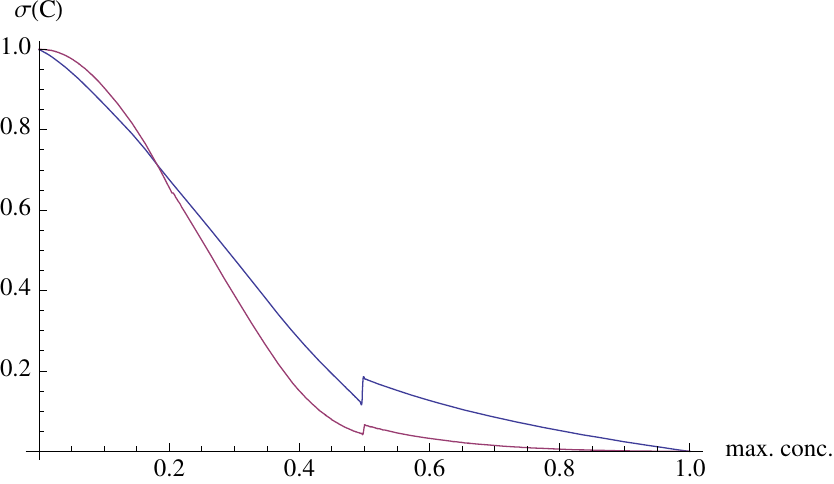}
\caption{\label{fig:Joint}
Joint plot of estimated (real [blue, $\beta=1$] and complex [red, $\beta=2$]) 
functions of the maximal concurrence over
spectral orbits, 
$S_4^{(1,\beta)}(\lambda_1\ldots\lambda_4) = \sigma^{(\beta)}
(C(\lambda\ldots\lambda_4))$. Note evident jumps
in both functions when the maximal concurrence equals 0.5. 
The graphs cross at $C=0.181245$. 
For $C = 0$, 
$\sigma(C)=1$, so all associated density matrices are separable.}
\end{figure}

We do find encouragement in advancing the ansatz 
(\ref{ansatz}) by the 
extensive numerical results we generate, in that our estimates
of $\sigma^{(1)}(C)$ and $\sigma^{(2)}(C)$ shown in Fig.~\ref{fig:Joint}
rather closely reproduce--as we will indicate below 
(sec.~\ref{evaluation})--other (independent)
numerical results and accompanying conjectures we have previously obtained.

The $\beta=2$ complex 
curve in Fig.~\ref{fig:Joint} shown is 
based on the use for quasi-Monte Carlo numerical integration of 
26,300,000 12-dimensional [Tezuka-Faure (TF) \cite{tezuka}] 
{\it low-discrepancy}
points, and the $\beta=1$ case, on 33,000,000 6-dimensional such 
points. (The TF procedure--programmed in Mathematica by Giray \"Okten 
\cite{giray1}--is not conducive to the placing of error bars
on the results, though later routines developed by him 
are.) These points comprise {\it sample} values, respectively, of 
the 12 Euler angles used to parameterize
$SU(4)$ and the 6 Euler angles used for $SO(4)$. For {\it each} 
TF-point, 499 auxiliary computations were carried out--in addition to that of 
the corresponding Haar measure associated with the Euler angles--for 
sets of eigenvalues
with values of maximal concurrence running at equally-spaced intervals 
from $\frac{1}{500}$ to $\frac{499}{500}$. 

Each density matrix generated--corresponding to a specific set of 
eigenvalues with fixed $C$ and Euler angles 
\cite{tbs} \cite[App. I]{slaterJMP2008}--was 
tested for separability. 
Prior to the quasi-Monte Carlo runs, we established
a database--using the Mathematica command ``FindInstance''--of 100 sets 
of four eigenvalues for each of the equally-spaced 499 values 
of $C$. One of the 100 sets 
was randomly selected [and then randomly permuted] for each of the TF-points and each of the 499 iterations. This ``random generation'' of 
sets of eigenvalues with {\it fixed} values of $C$ 
is clearly less than an ideal procedure, but it was what we found to be 
practical under the 
particular circumstances. (In sec.~\ref{supplementary}, 
we manage to improve upon this approach.)

Several weeks of MacMini 
computer time were used for 
each of the two sets--real and complex--of calculations. 
(Along with the computations concerning the maximal concurrence (\ref{maxcon}), we also carried out a fully parallel set of computations using the
related variable, 
$\frac{2 \sqrt{\lambda_2 \lambda_4}}{\lambda_1-\lambda_3}$. 
Those results, however, seemed 
comparatively disappointing in their predictive power, so
we do not detail them here.)
\subsection{Evaluation of numerical results} \label{evaluation}
Let us now appraise our estimated functions (Fig.~\ref{fig:Joint}) 
by seeing how well they are able to reproduce
previous related results, themselves based on very extensive analyses 
(mostly involving quasi-Monte Carlo integration also).
\subsubsection{Complex case}
Use of the complex ($\beta=2$) function in Fig.~\ref{fig:Joint} 
impressively explains
$98.7253\%$ of the variance of the estimated trivariate 
eigenvalue-parameterized separability function for $C>0$
presented in \cite[sec. III.B]{slaterJGP2}. It also yields an estimate
of 0.254756 for the Hilbert-Schmidt separability probability, while our exact
conjecture from \cite{slater833} is $\frac{8}{33} \approx 0.242424$.
Further, the {\it Bures} separability probability estimate yielded is 
0.0692753,
while our conjectured value is $\frac{1680 \left(-1+\sqrt{2}\right)}{\pi ^8} 
\approx 0.0733389$ \cite{slaterJGP}.
\subsubsection{Real case}
Passing to the real ($\beta=1$) case, we had previously formulated 
the conjecture that
the HS separability probability is $\frac{8}{17} \approx 0.470588$ 
\cite[sec. 9.1]{slater833}.
Our estimate based on the (blue) function shown in Fig.~\ref{fig:Joint}
is 0.480302. (The corresponding estimate--for which we have no prior
conjecture--for the Bures {\it real} two-qubit 
separability probability is 0.212152.)
Further, we are able to reproduce $97.7502\%$ of the variation in the
corresponding trivariate function for $C>0$. 
(This last function had been estimated using
a recent Euler-angle parameterization of $SO(4)$, obtained by S. Cacciatori
\cite[App. I]{slaterJMP2008}. It was derived by Cacciatori 
after the submission of \cite{slaterJGP2}, 
and thus not reported nor used 
there, although its complex counterpart--based 
on 3,600,000 
Tezuka-Faure points--had been
\cite[sec. III.B]{slaterJGP2},
while the real case was based on a considerably lesser number of 
TF-points, 700,000.)
\subsection{Jumps near $C=\frac{1}{2}$}
\begin{figure}
\includegraphics{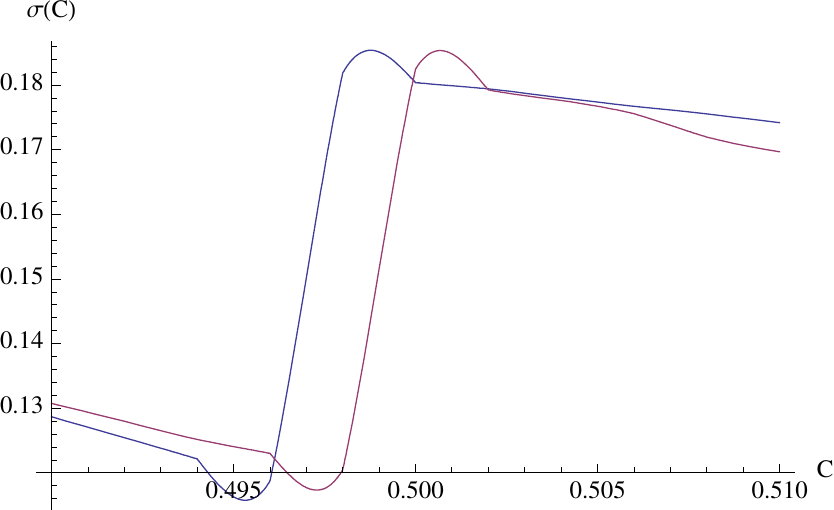}
\caption{\label{fig:midpoint}Same plot as Fig.~\ref{fig:Joint},
restricted to the vicinity of $C=\frac{1}{2}$, and the complex (red) 
curve being amplified by a factor of 2.8}
\end{figure}

For the real ($\beta=1$) case, the 
jump near $C(\lambda_1\ldots\lambda_4)=\frac{1}{2}$ is from approximately 
0.118696 to 0.180357, and in the complex ($\beta=2$) case,
from 0.0439255 to 0.651586. 
The magnitudes of the two jumps are then quite 
comparable, being respectively,
$51.964\%$ and $51.488\%$. 
In Fig.~\ref{fig:midpoint}, we replot the curves shown in Fig.~\ref{fig:Joint}
in the immediate vicinity of $C=\frac{1}{2}$, but amplify the complex 
(red) curve by a factor of 2.8. We perceive a very close similarity
in shape.
\subsection{Additional discontinuities}
\begin{figure}
\includegraphics{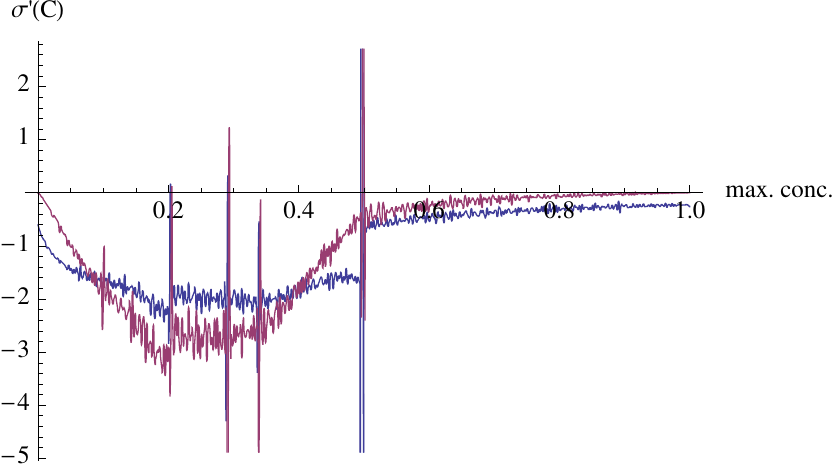}
\caption{\label{fig:Diff}
Joint plot of {\it derivatives} with respect to $C$ of the 
estimated (real [blue, $\beta=1$] and complex [red, $\beta=2$])
functions of the maximal concurrence over
spectral orbits,
$\sigma^{(\beta)}(C(\lambda\ldots\lambda_4))$. 
Spikes are observable--{\it both} for the real and complex 
cases--at: $\frac{1}{2}=0.5;
\frac{147}{500} =0.294$; $\frac{51}{250}=0.204$; and
$\frac{17}{50} =\frac{51}{150}
=0.34$.}
\end{figure}

\begin{figure}
\includegraphics{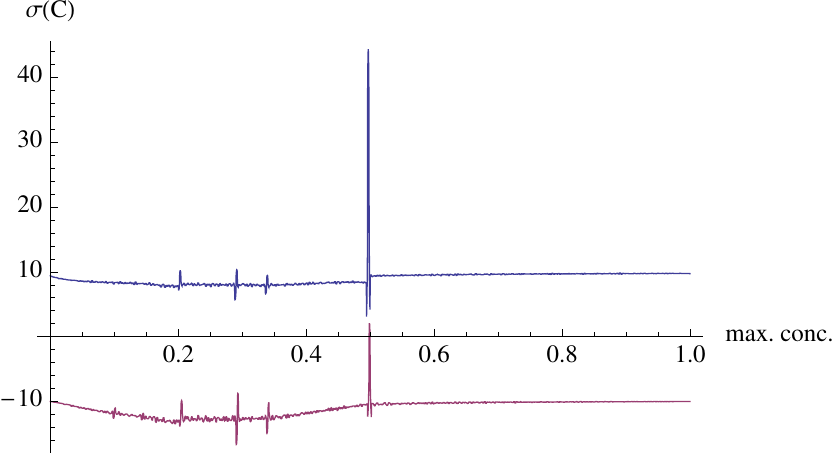}
\caption{\label{fig:Diff2}Same as Fig.~\ref{fig:Diff}, except that the real
(blue) curve has been translated upwards by 10 units and the complex (red) 
curve downwards by 10 units, so that the individual 
discontinuities in the two derivatives can be more readily seen 
and compared.}
\end{figure}

In Fig.~\ref{fig:Diff} we show the {\it derivatives}
with respect to $C$ of the estimates of $\sigma^{(\beta)}(C)$. (Fig.~\ref{fig:Diff2} is a plot of the same two curves, except that we have now 
added 10 to the derivative in the 
real case and subtracted 10 in the complex case, 
so that the discontinuities can be more readily distinguished and compared.)
In addition to the already-discussed behavior at $C={1/2}$, we see--{\it both} in the real and complex cases--a 
secondary spike at $\frac{147}{500} =0.294$, and 
lesser spikes at $\frac{51}{250}=0.204$ and $\frac{17}{50} =\frac{51}{150} 
=0.34$.
So, all the observed spikes, signaling what we presume are discontinuities
in the $\sigma^{(\beta)}$'s, and concomitant 
nontrivial {\it piecewise} behavior--indicative of different separability
constraints becoming active/binding or not--are 
for $C \leq \frac{1}{2}$. The point $C=\frac{51}{500}=0.102$ may also
be a discontinuity, at least in the complex case.
We could detect no apparent spikes/discontinuities in the upper half-range, 
$C \in [\frac{1}{2},1]$. (In a 
somewhat analogous study of two-qubit {\it three}-parameter 
HS separability {\it probabilities}, intricate {\it piecewise} 
continuous behavior
[involving the {\it golden ratio}] was observed \cite[eq. (37) 
and Fig. 11]{pbsCanosa}.)

\begin{figure}
\includegraphics{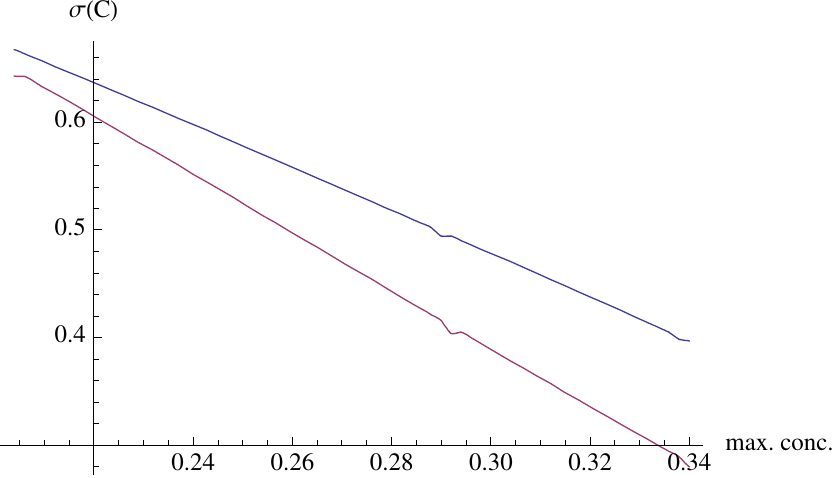}
\caption{\label{fig:segment}
Joint plot of estimated (real [blue, $\beta=1$] and complex [red, $\beta=2$])
functions of the maximal concurrence over
spectral orbits,
$S_4^{(1,\beta)}(\lambda_1\ldots\lambda_4) = \sigma^{(\beta)}
(C(\lambda\ldots\lambda_4))$. The graphs are 
obviously close to linear between
the discontinuities $\frac{51}{250}=0.204$
and $\frac{17}{50} =0.34$, 
except for the intermediate 
discontinuity at $\frac{147}{500} =0.294$. 
To high accuracy, the real (blue) curve can be fitted by the line
$1.07614 -1.99472C$ and the complex (red) curve, by
$1.19822 - 2.69548 C$.}
\end{figure}

In Fig.~\ref{fig:segment} we show the segments of the estimated functions
$\sigma^{(\beta)}(C)$ between the two discontinuities, $\frac{51}{250}=0.204$ 
and $\frac{17}{50} =0.34$. The behavior seems very close to {\it linear} 
for both curves,
except for the intermediate discontinuity at $\frac{147}{500} =0.294$.
\subsection{Supplementary analyses} \label{supplementary}
\begin{figure}
\includegraphics{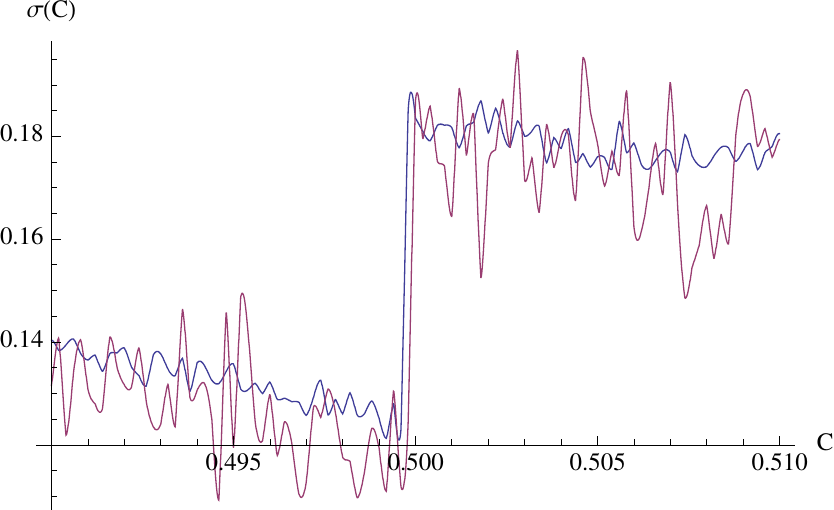}
\caption{\label{fig:midpoint2}The same plot as Fig.~\ref{fig:midpoint},
but based on our ongoing supplementary analysis with finer resolution 
in $C$
and enhanced eigenvalue sampling. The twin jumps 
in the estimated eigenvalue-parameterized real and complex 
separability functions near $C=\frac{1}{2}$ are now 
certainly indisputably clear.}
\end{figure}

\begin{figure}
\includegraphics{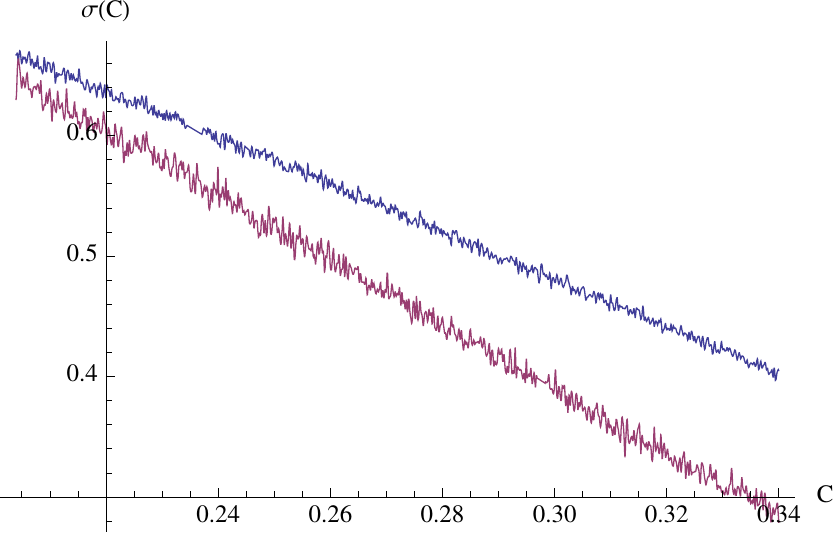}
\caption{\label{fig:segment2}Same plot as Fig.~\ref{fig:segment}, but 
based on our ongoing supplementary analysis with finer resolution in $C$ 
and enhanced eigenvalue sampling}
\end{figure}

Since the completion of the extensive numerical analyses described above,
we have undertaken supplementary, parallel analyses in which 5,000 
(rather than 500) subintervals
of $C \in [0,1]$ are employed, as well as an improved method is used
for sampling random eigenvalues with fixed values of $C$ (using the 
Mathematica FindInstance command, now with a random seed). These results 
so far seem largely
consistent with those already described.
However, the new analogues of
the plots of derivatives,
Figs.~\ref{fig:Diff} and \ref{fig:Diff2},
are still much too rough in character
to detect the presence of any secondary (non-jump) discontinuities.
But, even at this stage (having tested 
30,400 times the separability of 4,999 complex density matrices, 
and 28,100  times the separability of 4,999 real density matrices),
we can produce the interesting counterpart 
(Fig.~\ref{fig:midpoint2}) to Fig.~\ref{fig:midpoint}, in which
the two jumps near $C=\frac{1}{2}$ of roughly $50\%$ 
magnitude are clearly unmistakable.
Further, the highly linear behavior
displayed in Fig.~\ref{fig:segment} also reappears (Fig.~\ref{fig:segment2}).

\section{Concluding Remarks}
For the real and complex two-qubit systems, we have 
investigated the possibility
that the associated ({\it
three}-dimensional) {\it eigenvalue}-parameterized
separability
functions--conceptually important for computing separability
{\it probabilities}--are
expressible as {\it one}-dimensional functions ($\sigma(C)$)
of the maximal concurrence over spectral orbits ($C \in [0,1]$).
Our numerical estimates, in this regard,
have been encouraging, in that they closely reproduce
{\it independently}-generated
numerical results and exact conjectures concerning
separability probabilities based on the Hilbert-Schmidt and Bures
(minimal monotone)
metrics over the two-qubit systems, and based on
the use of {\it diagonal-entry}-parameterized
separability functions. Plots of the real and complex versions of
$\sigma(C)$ {\it both}
exhibit {\it jumps} of approximately $50\%$ magnitude
near the midpoint, $C=\frac{1}{2}$, and {\it both}
also indicate the presence of,
at least, three further (non-jump)
discontinuities ($C \approx 0.204, 0.294, 0.34$), apparently indicative
of points at which certain distinct
separability constraints become either
active/binding or not.
Over the interval $C \in [0.204,0.34]$, the real and complex fitted
functions $\sigma(C)$ {\it both} appear to be
simply linear (except at $C \approx 0.294$).

We have principally studied above the possibility (4)
that the ostensibly {\it trivariate}
two-qubit eigenvalue-separability functions can be 
equivalently expressed as {\it univariate}
functions of only a single variable, that is, the maximal concurrence $C$ 
over spectral orbits \cite{roland2}. 
Since we have unfortunately 
not been able to fully formally
resolve this issue--although our supporting evidence
for this proposition is 
intriguing--we can not also presently fully eliminate the possibility 
that one or even two (yet unspecified) 
variables supplemental to $C$ are, in fact, needed, 
and that the corresponding separability function is not in fact strictly
univariate in nature (as we do know it definitely is the case with 
the two-qubit diagonal-entry-parameterized separability functions 
\cite{slater833}).

It presently 
appears somewhat problematical to extend the line of analysis above
to the qubit-{\it qutrit} ($N=6$) case. In addition, to simply the 
greatly increased
computational burden that would be involved, there does not seem to be a 
maximal concurrence formula comparable to 
the two-qubit one 
(\ref{maxcon}) with the requisite properties 
we have utilized \cite[p. 102108-16]{roland2}.

We have examined the relationship between separability
and entanglement in a specific analytical setting--involving 
{\it eigenvalue}-parameterized 
separability functions and the use of the \.Zyczkowski-Sommers formulas
\cite{szHS,szBures} for the total Hilbert-Schmidt and Bures volumes of the $N \times N$ 
density matrices. A number of studies of Batle, Casas, A. Plastino and A. R. 
Plastino 
(for example, \cite{batlecasas1})
have also focused on the relationship between separability and entanglement, 
but in somewhat different analytical frameworks (typically involving 
the ZHSL measure \cite{ZHSL}, which is {\it uniform} over 
the eigenvalue simplex). 
 The closest we can come, it seems, to a direct comparison 
with their analyses,
is to note that the dot-dashed curve in Fig. 2 
of \cite{batlecasas1} is based
on the Hilbert-Schmidt metric, and their $x$-axis is the Bures distance,
while we have employed the maximal concurrence $C$ on the $x$-axis in 
the somewhat comparable Fig. 1 above (cf. \cite{batleplastino1}). 
Both theirs and our plots are, in general, downward-decreasing, but
theirs gives no indication of any discontinuities.
Also, their plot is of the separability {\it probability}, while ours
is of the (presumed univariate) 
eigenvalue-parameterized 
separability {\it function}, to be used in the computation of the probability.

\begin{acknowledgments}
I would like to express appreciation to the Kavli Institute for Theoretical
Physics (KITP)
for computational support in this research.
\end{acknowledgments}

\bibliography{MaxConcur4}

\end{document}